\documentstyle[12pt]{article}

\catcode`\@=11
\long\def\@makefntext#1{ %\parindent 1em
\protect\noindent \hbox to 3.2pt {\hskip-.9pt
$^{{\ninerm\@thefnmark}}$\hfil}#1\hfill} %can be used

\def\thefootnote{\fnsymbol{footnote}}
 \def\@makefnmark{\hbox to 0pt{$^{\@thefnmark}$\hss}}  %original

\def\ps@myheadings{\let\@mkboth\@gobbletwo
\def\@oddhead{\hbox{} %\sl
\rightmark\hfil\ninerm\thepage}
\def\@oddfoot{}\def\@evenhead{\ninerm\thepage\hfil %\sl
\leftmark\hbox{}}\def\@evenfoot{}
\def\sectionmark##1{}\def\subsectionmark##1{}}

\newcommand{\beq}{\begin{equation}}
\newcommand{\eeq}{\end{equation}}
\newcommand{\bea}{\begin{eqnarray}}
\newcommand{\eea}{\end{eqnarray}}
\newcommand{\barr}{\begin{array}}
\newcommand{\earr}{\end{array}}
\newcommand{\bc}{\begin{center}}
\newcommand{\ec}{\end{center}}
\newcommand{\btab}{\begin{tabular}}
\newcommand{\etab}{\end{tabular}}
\newcommand{\gv}{\mbox{GeV}}

\newcommand{\mv}{\mbox{MeV}}
\newcommand{\kv}{\mbox{keV}}
\newcommand{\ev}{\mbox{eV}}
\newcommand{\nn}{\nonumber}
\newcommand{\ra}{\rightarrow}

\newcommand{\sz}{\Sigma^{ZZ}}
\newcommand{\sw}{\Sigma^{WW}}

\newcommand{\dro}{\Delta\rho}
\newcommand{\drqcd}{\delta\!\rho\,_{QCD}}
\newcommand{\roro}{\rho^{(2)}}

\newcommand{\drb}{\Delta\overline{\rho}}

\newcommand{\al}{\alpha}

\newcommand{\g}{\gamma}
\newcommand{\G}{\Gamma}
\newcommand{\Gmu}{G_{\mu}}
\newcommand{\Gtau}{G_{\tau}}

\newcommand{\ganu}{\gamma_{\nu}}

\newcommand{\gafi}{\gamma_5}

\newcommand{\Pig}{\Pi^{\gamma}}

\newcommand{\noi}{\noindent}
\newcommand{\epmf}{e^+e^- \rightarrow f\bar{f}}

\newcommand{\sm}{Standard Model }
\newcommand{\su}{SU(2)$\times$U(1) }

\newcommand{\dal}{\Delta\alpha}

\newcommand{\mz}{M_Z^2}
\newcommand{\mw}{M_W^2}

\newcommand{\real}{\mbox{Re}}

\newcommand{\Dr}{\Delta r}

\newcommand{\eps}{\epsilon}

\newcommand{\aspi}{\frac{\alpha_s}{\pi}}

\newcommand{\ass}{asymmetries }

\newcommand{\pr}{Phys.\ Rev.\ }
 \newcommand{\prd}{Phys.\ Rev.\ D }
\newcommand{\zp}{Z.\ Phys.\ C }
\newcommand{\plb}{Phys.\ Lett.\ B }
 \newcommand{\prl}{Phys.\ Rev.\ Lett.\ }
\newcommand{\np}{Nucl.\ Phys.\ B }

\newcommand{\ms}{\overline{MS}}

\begin{document}

%----------------------------PROCSLA.STY---------------------------------------
\newcommand{\symbolfootnote}{\renewcommand{\thefootnote}
        {\fnsymbol{footnote}}}
\renewcommand{\thefootnote}{\fnsymbol{footnote}}
\newcommand{\alphfootnote}
        {\setcounter{footnote}{0}
         \renewcommand{\thefootnote}{\sevenrm\alph{footnote}}}

%------------------------------------------------------------------------------
%NEW DEFINED SECTION COMMANDS
\newcounter{sectionc}\newcounter{subsectionc}\newcounter{subsubsectionc}
\renewcommand{\section}[1] {\vspace{0.6cm}\addtocounter{sectionc}{1}
\setcounter{subsectionc}{0}\setcounter{subsubsectionc}{0}\noindent
        {\bf\thesectionc. #1}\par\vspace{0.4cm}}
\renewcommand{\subsection}[1] {\vspace{0.6cm}\addtocounter{subsectionc}{1}
        \setcounter{subsubsectionc}{0}\noindent
        {\it\thesectionc.\thesubsectionc. #1}\par\vspace{0.4cm}}
\renewcommand{\subsubsection}[1]
{\vspace{0.6cm}\addtocounter{subsubsectionc}{1}
        \noindent {\rm\thesectionc.\thesubsectionc.\thesubsubsectionc.
        #1}\par\vspace{0.4cm}}
\newcommand{\nonumsection}[1] {\vspace{0.6cm}\noindent{\bf #1}
        \par\vspace{0.4cm}}

%NEW MACRO TO HANDLE APPENDICES
\newcounter{appendixc}
\newcounter{subappendixc}[appendixc]
\newcounter{subsubappendixc}[subappendixc]
\renewcommand{\thesubappendixc}{\Alph{appendixc}.\arabic{subappendixc}}
\renewcommand{\thesubsubappendixc}
        {\Alph{appendixc}.\arabic{subappendixc}.\arabic{subsubappendixc}}

\renewcommand{\appendix}[1] {\vspace{0.6cm}
        \refstepcounter{appendixc}
        \setcounter{figure}{0}
        \setcounter{table}{0}
        \setcounter{equation}{0}
        \renewcommand{\thefigure}{\Alph{appendixc}.\arabic{figure}}
        \renewcommand{\thetable}{\Alph{appendixc}.\arabic{table}}
        \renewcommand{\theappendixc}{\Alph{appendixc}}
        \renewcommand{\theequation}{\Alph{appendixc}.\arabic{equation}}
%       \noindent{\bf Appendix \theappendixc. #1}\par\vspace{0.4cm}}
        \noindent{\bf Appendix \theappendixc #1}\par\vspace{0.4cm}}
\newcommand{\subappendix}[1] {\vspace{0.6cm}
        \refstepcounter{subappendixc}
        \noindent{\bf Appendix \thesubappendixc. #1}\par\vspace{0.4cm}}
\newcommand{\subsubappendix}[1] {\vspace{0.6cm}
        \refstepcounter{subsubappendixc}
        \noindent{\it Appendix \thesubsubappendixc. #1}
        \par\vspace{0.4cm}}

%------------------------------------------------------------------------------
%MARCO FOR ABSTRACT BLOCK
\def\abstracts#1{{
        \centering{\begin{minipage}{30pc}\tenrm\baselineskip=12pt\noindent
        \centerline{\tenrm ABSTRACT}\vspace{0.3cm}
        \parindent=0pt #1
        \end{minipage} }\par}}

%------------------------------------------------------------------------------
%NEW MACRO FOR BIBLIOGRAPHY
\newcommand{\bibit}{\it}
\newcommand{\bibbf}{\bf}
\renewenvironment{thebibliography}[1]
        {\begin{list}{\arabic{enumi}.}
        {\usecounter{enumi}\setlength{\parsep}{0pt}
%1.25cm IS STRICTLY FOR PROCSLA.TEX ONLY
\setlength{\leftmargin 1.25cm}{\rightmargin 0pt}
%0.52cm IS FOR NEW DATA FILES
%\setlength{\leftmargin 0.52cm}{\rightmargin 0pt}
         \setlength{\itemsep}{0pt} \settowidth
        {\labelwidth}{#1.}\sloppy}}{\end{list}}

%------------------------------------------------------------------------------
%FOLLOWING THREE COMMANDS ARE FOR 'LIST' COMMAND.
\topsep=0in\parsep=0in\itemsep=0in
\parindent=1.5pc

%LIST ENVIRONMENTS
\newcounter{itemlistc}
\newcounter{romanlistc}
\newcounter{alphlistc}
\newcounter{arabiclistc}
\newenvironment{itemlist}
        {\setcounter{itemlistc}{0}
         \begin{list}{$\bullet$}
        {\usecounter{itemlistc}
         \setlength{\parsep}{0pt}
         \setlength{\itemsep}{0pt}}}{\end{list}}

\newenvironment{romanlist}
        {\setcounter{romanlistc}{0}
         \begin{list}{$($\roman{romanlistc}$)$}
        {\usecounter{romanlistc}
         \setlength{\parsep}{0pt}
         \setlength{\itemsep}{0pt}}}{\end{list}}

\newenvironment{alphlist}
        {\setcounter{alphlistc}{0}
         \begin{list}{$($\alph{alphlistc}$)$}
        {\usecounter{alphlistc}
         \setlength{\parsep}{0pt}
         \setlength{\itemsep}{0pt}}}{\end{list}}

\newenvironment{arabiclist}
        {\setcounter{arabiclistc}{0}
         \begin{list}{\arabic{arabiclistc}}
        {\usecounter{arabiclistc}
         \setlength{\parsep}{0pt}
         \setlength{\itemsep}{0pt}}}{\end{list}}

%------------------------------------------------------------------------------
%FIGURE CAPTION
\newcommand{\fcaption}[1]{
        \refstepcounter{figure}
        \setbox\@tempboxa = \hbox{\tenrm Fig.~\thefigure. #1}
        \ifdim \wd\@tempboxa > 6in
           {\begin{center}
        \parbox{6in}{\tenrm\baselineskip=12pt Fig.~\thefigure. #1 }
            \end{center}}
        \else
             {\begin{center}
             {\tenrm Fig.~\thefigure. #1}
              \end{center}}
        \fi}

%TABLE CAPTION
\newcommand{\tcaption}[1]{
        \refstepcounter{table}
        \setbox\@tempboxa = \hbox{\tenrm Table~\thetable. #1}
        \ifdim \wd\@tempboxa > 6in
           {\begin{center}
        \parbox{6in}{\tenrm\baselineskip=12pt Table~\thetable. #1 }
            \end{center}}
        \else
             {\begin{center}
             {\tenrm Table~\thetable. #1}
              \end{center}}
        \fi}

%------------------------------------------------------------------------------
%ACKNOWLEDGEMENT: this portion is from John Hershberger
\def\@citex[#1]#2{\if@filesw\immediate\write\@auxout
        {\string\citation{#2}}\fi
\def\@citea{}\@cite{\@for\@citeb:=#2\do
        {\@citea\def\@citea{,}\@ifundefined
        {b@\@citeb}{{\bf ?}\@warning
        {Citation `\@citeb' on page \thepage \space undefined}}
        {\csname b@\@citeb\endcsname}}}{#1}}

\newif\if@cghi
\def\cite{\@cghitrue\@ifnextchar [{\@tempswatrue
        \@citex}{\@tempswafalse\@citex[]}}
\def\citelow{\@cghifalse\@ifnextchar [{\@tempswatrue
        \@citex}{\@tempswafalse\@citex[]}}
\def\@cite#1#2{{$\null^{#1}$\if@tempswa\typeout
        {IJCGA warning: optional citation argument
        ignored: `#2'} \fi}}
\newcommand{\citeup}{\cite}

%------------------------------------------------------------------------------
%FOR FNSYMBOL FOOTNOTE AND ALPH{FOOTNOTE}
\def\fnm#1{$^{\mbox{\scriptsize #1}}$}
\def\fnt#1#2{\footnotetext{\kern-.3em
        {$^{\mbox{\sevenrm #1}}$}{#2}}}

%------------------------------------------------------------------------------
\font\twelvebf=cmbx10 scaled\magstep 1
\font\twelverm=cmr10 scaled\magstep 1
\font\twelveit=cmti10 scaled\magstep 1
\font\elevenbfit=cmbxti10 scaled\magstephalf
\font\elevenbf=cmbx10 scaled\magstephalf
\font\elevenrm=cmr10 scaled\magstephalf
\font\elevenit=cmti10 scaled\magstephalf
\font\bfit=cmbxti10
\font\tenbf=cmbx10
\font\tenrm=cmr10
\font\tenit=cmti10
\font\ninebf=cmbx9
\font\ninerm=cmr9
\font\nineit=cmti9
\font\eightbf=cmbx8
\font\eightrm=cmr8
\font\eightit=cmti8

%----------------------START OF DATA FILE------------------------------

\centerline{\tenbf PRECISION TESTS OF THE ELECTROWEAK INTERACTION
%\baselineskip=16pt
%\centerline{\tenbf PHENOMENOLOGY
                    \footnote{invited talk at the
                    XV Int. Conference on
                    Physics in Collisions,
                    Cracow, Poland, June 1995} }
%\baselineskip=22pt
%\centerline{\tenbf INSTRUCTIONS FOR TYPESETTING CAMERA-READY}
%\baselineskip=16pt
%\centerline{\tenbf MANUSCRIPT USING COMPUTER SOFTWARE}
%\centerline{\ninerm (For 20\% Reduction to 6 in. $\times$ 8.5 in. Trim Size)}
\vspace{0.8cm}
\centerline{\tenrm WOLFGANG HOLLIK }
%                   \footnote{supported in part by the European Union
%                    under contract CHRX-CT92-0004} }
\baselineskip=13pt
\centerline{\tenit Institut f\"ur Theoretische Physik,
                   Universit\"at Karlsruhe}
\baselineskip=12pt
\centerline{\tenit D-76128 Karlsruhe, Germany} \hfill \\[0.5cm]
%\vspace{0.9cm}
\abstracts{The status of the electroweak
Standard Model is reviewed in the light of
recent precision data and new
theoretical results which have
contributed to improve the predictions for
precision observables, together with the remaining inherent theoretical
uncertainties. Consequences for possible new physics are also
discussed.}

\vspace*{0.5cm}
%\vfil
%\vspace{0.8cm}
\twelverm   %modified by CLee 23/07/93
\baselineskip=14pt
\section{Standard Model entries:}
\vspace*{-0.7cm}
\subsection{The fermions}
\vspace*{-0.35cm}
%\vglue 0.4cm
%\vglue 0.3cm
%\leftline{\twelveit 1.1. The fermion families}
%\vglue 0.4cm
%\vglue 1pt
The family structure of the fermions is a manifestation of the
\su symmetry. It has been strongly  consolidated
by several recent experimental informations:

\smallskip \noi
{\it Three generations of massless neutrinos:}
{}From the measurements of the $Z$ line shape at LEP
the combined LEP value for the number of light neutrinos
is \cite{LEP}
(universal couplings assumed)
   $$ N_{\nu} = 2.987\pm 0.017 \, . $$
$m_{\nu} = 0 $ is consistent with the experimental mass limits
experiments \cite{smirnov}
$$
 m_{\nu_e} < 7.2\, \ev \, (95\% \, C.L.), \;\;\;
 m_{\nu_{\mu}} < 220\, \kv \, (90\%\, C.L.), \;\;\;
 m_{\nu_{\tau}} < 24\,  \mv \, (95\%\, C.L.)  \, .  $$

\smallskip \noi
{\it Universality of neutral current couplings:}
The vector and axial vector coupling constants of the $Z$ to
$e,\mu,\tau$ measured at LEP \cite{LEP}
 show agreement with lepton universality
and with the \sm prediction  (Figure 1).

Recent results on
 $\sigma(\nu_{\mu} e)$ and
 $\sigma(\bar{\nu}_{\mu} e)$
 by the CHARM II Collaboration  yield for the
        $\nu_{\mu}$ and $e$ coupling constants
 \cite{beyer}
 \bea
&   & g_V^{\nu e} \equiv 2g^{\nu} \, g_V^e
 = -0.035\pm 0.017 \nn \\
  &   & g_A^{\nu e} \equiv 2g^{\nu} \, g_A^e = -0.503
        \pm 0.017 , \nn
\eea
compatible with $g_{V,A}^{lept}$ from LEP \cite{LEP} under the assumption
of lepton universality:
\bea
      g^{\ell}_V & = & -0.0366 \pm 0.0011 \nn \\
      g^{\ell}_A & = & - 0.50123 \pm 0.00042 \, . \nn
\eea

\begin{figure} % fig 1
\vspace*{3.25in}
\caption{
            68\% C.L. contours for the leptonic coupling constants
            from LEP, ref (1).}
\end{figure}

\smallskip \noi
{\it Universality of charged current couplings:}
The $\tau$-$\mu$ CC universality can be expressed in terms of
the ratio of the effective decay constants $\Gtau$ for
$\tau \ra \nu_{\tau} e \bar{\nu}_e$ and $\Gmu$ for
$\mu  \ra \nu_{\mu } e \bar{\nu}_e$ to be unity in the
Standard Model:
$$
 \left( \frac{\Gtau}{\Gmu}\right)^2 = B_e\cdot
 \frac{\tau_{\mu}}{\tau_{\tau}} \left(
 \frac{m_{\mu}}{m_{\tau}} \right)^5 = 1 \, .
$$
The recent data on the $\tau$ mass $m_{\tau}$, the $\tau$ lifetime
$\tau_{\tau}$, and the branching ratio
$B_e = BR(\tau \ra \nu_{\tau} e \bar{\nu}_e)$ yield
\cite{schwarz}
\beq
 \left( \frac{\Gtau}{\Gmu}\right)^2 = 0.996 \pm 0.006 \, ,
\eeq
consistent with CC $\tau$-$\mu$ universality.
The CC $\mu$-$e$ universality is demonstrated in terms of the
experimental ratios \cite{schwarz}
\bea
 B_e  = BR(\tau \ra \nu_{\tau} e \bar{\nu}_e)
 & = & 0.1789\pm 0.0014 \nn \\
 B_{\mu} = BR(\tau \ra \nu_{\tau} \mu \bar{\nu}_{\mu})
 & = & 0.1734 \pm 0.0016 \, . \nn
\eea
By purely kinematical reasons, $B_{\mu} = 0.972 B_e$,
which actually is observed in the experimental ratios of Eq.\ (2).

\smallskip \noi
{\it The top quark:}
The top quark has recently been observed at the Tevatron. Its mass
determination by the CDF collaboration \cite{top} yields
$m_t = 176 \pm 8 \pm 10$ GeV and by the D0 collaboration
 \cite{d0}:
$m_t = 199^{+19}_{-21} \pm 22$ GeV, resulting in an weighted average of
\beq            m_t = 180 \pm 12 \gv \, . \eeq
%
%\subsection{The vector bosons and the Higgs sector}
%\vglue 0.4cm
\vglue 0.3cm
\leftline{\twelveit 1.2. The vector bosons and the Higgs sector}
%\vglue 0.4cm
\vglue 1pt
The spectrum of the vector bosons $\g,W^{\pm},Z$  with
masses       \cite{LEP,pp}
 \beq
  M_W = 80.26\pm 0.16\, \gv, \;\;\;\;
  M_Z = 91.1887\pm 0.0022\, \gv
\eeq
is reconciled with the \su local gauge symmetry with the help of the
Higgs mechanism. For a general structure of the scalar sector,
the electroweak mixing angle is related to the vector boson
masses  by
\beq
 s^2_{\theta} \equiv \sin^2\theta =
 1-\frac{\mw}{\rho\mz} =
   1-\frac{\mw}{\mz} + \frac{\mw}{\mz} \dro
 \equiv s_W^2 + c_W^2 \dro
\eeq
where the $\rho$-parameter  $\rho = (1-\dro)^{-1}$
is an additional free parameter.
In models with scalar doublets only, in particular in the
minimal model, one has the tree level relation $\rho=1$.
Loop effects, however, induce a deviation $\dro \neq 0$.

We can get a value
for $\rho$ from directly using the data on $M_W,M_Z$
and the mixing angle
    $s^2_\theta = s_{\ell}^2 = 0.2318 \pm0.0004$
measured at LEP
\cite{LEP}
as independent experimental information,
 yielding
$
   \rho   =   \mw/\mz c_{\ell}^2
      = 1.0084 \pm 0.0041 \, .
$
In the \sm $M_W,M_Z,s^2_{\ell}$ are correlated.
Taking into account the constraints from the data
\cite{LEP}
yields      $\rho_{SM} = 1.0066\pm 0.0010$.
The deviation $\rho-\rho_{SM}$
 can be interpreted as a
measure for a possibly deviating tree level structure. The present situation
is consistent with the Standard Model.

\section{Precision tests of the Standard Model}
\vspace*{-0.7cm}
\subsection{Loop calculations}
\vspace*{-0.35cm}

The possibility of performing precision tests is based
on the formulation of the \sm as a renormalizable quantum field
theory preserving its predictive power beyond tree level
calculations. With the experimental accuracy in the investigation
of the fermion-gauge boson interactions being sensitive to the loop
induced quantum effects, also the more subtle parts of the \sm
Lagrangian are probed.

Before one can make predictions from the theory,
a set of independent parameters has to be determined from experiment.
All the practical schemes make use of the same physical input quantities
\beq \al, \; \Gmu,\; M_Z,\; m_f,\; M_H \eeq
for fixing the free parameters of the SM.
 Differences between various schemes are formally
of higher order than the one under consideration.
 The study of the
scheme dependence of the perturbative results, after improvement by
resumming the leading terms, allows us to estimate the missing
higher order contributions.

%\smallskip \noi
\newpage \noi
{\it Large loop effects in electroweak parameter shifts:}
\begin{itemize}
\item[(i)]
The fermionic content of the subtracted photon vacuum polarization
$$
 \dal =   \Pig_{ferm}(0) -
     \real\,\Pig_{ferm}(\mz)
$$
corresponds to a QED induced shift
in the electromagnetic fine structure constant. The recent update of the
evaluation of the light quark content
by Eidelman and Jegerlehner \cite{eidelman} and Burkhardt and Pietrzyk
\cite{burkhardt} both yield the result
$$ (\dal)_{had} = 0.0280 \pm 0.0007 $$
and thus
confirm the previous value of \cite{vacpol} with an improved accuracy.
Other determinations by Swartz \cite{swartz} and Martin and Zeppenfeld
\cite{martin} agree within one standard deviation. Together with the leptonic
content, $\dal$ can
be resummed resulting in an effective fine structure
constant at the $Z$ mass scale:
\beq
   \al(\mz) \, =\, \frac{\al}{1-\dal}\,=\,
   \frac{1}{128.89\pm 0.09} \, .
\eeq
 \item[(ii)]
The $\rho$-parameter  in the Standard Model
gets a deviation $\dro$ from 1 by radiative corrections,  essentially
by  the contribution of the $(t,b)$ doublet \cite{rho},
 in 1-loop and
 neglecting $m_b$:
\beq   \left[
     \frac{\sz(0)}{\mz} -\frac{\sw(0)}{\mw}
       \right]_{(t,b)} \, = \,
        \frac{3\Gmu m_t^2}{8\pi^2\sqrt{2}}   \, = \, \dro \, .
\eeq
This potentially large
contribution constitutes also the leading
shift for the electroweak mixing angle
when inserted into Eq.\ (4).
\end{itemize}

%\subsection{The vector boson masses}
%\vglue 0.4cm
\vglue 0.3cm
\leftline{\twelveit 2.2. The vector boson masses}
%\vglue 0.4cm
\vglue 1pt
The correlation between
the masses $M_W,M_Z$ of the vector bosons          in terms
of the Fermi constant $\Gmu$       reads in 1-loop order
of the Standard Model \cite{sirmar}:
\beq
\frac{\Gmu}{\sqrt{2}}   =
            \frac{\pi\al}{2s_W^2 M_W^2} \left[
        1+ \Dr(\al,M_W,M_Z,M_H,m_t)\right]\, .
\eeq
The 1-loop correction $\Dr$ can be written in the following way
\beq
 \Dr = \Delta\al -\frac{c_W^2}{s_W^2}\,\dro
         + (\Dr)_{remainder} \, .
\eeq
in order to separate the
leading fermionic contributions
                $\dal$ and $\dro$.
All other terms are collected in
the $(\Dr)_{remainder}$,
the typical size of which is of the order $\sim 0.01$.

\bigskip
The presence of large terms in $\Dr$ requires the consideration
of higher than 1-loop effects.
The modification of Eq.\ (8) according to
\beq
         1+\Dr  \, \ra\, \frac{1}{(1-\Delta\al)\cdot
(1+\frac{c_W^2}{s_W^2}\drb) \, -\,(\Dr)_{remainder}}
 \equiv \frac{1}{1-\Dr}
\eeq
with
\beq
    \drb  =   3\,
 \frac{\Gmu m_t^2}{8\pi^2\sqrt{2}}\cdot \left[1+
 \frac{\Gmu m_t^2}{8\pi^2\sqrt{2}}\roro)
 + \drqcd  \right]
\eeq
accommodates the following higher order terms
($\Dr$ in the denominator is an effective correction including
higher orders):
\begin{itemize}
\item
The leading log resummation \cite{marciano} of $\dal$:
$  1+\dal\, \ra \, (1-\dal)^{-1}$
\item
The resummation of the leading $m_t^2$ contribution \cite{chj}
 in terms
of $\drb$. Thereby also irreducible higher order
diagrams contribute. The electroweak 2-loop
 part is described by the
function $\roro(M_H/m_t)$ derived in \cite{barbieri} for general
Higgs masses.
$\drqcd$ is the QCD correction
to the leading $m_t^2$ term of the $\rho$-parameter \cite{djouadi,tarasov}
\beq
    \drqcd = -\,
\frac{\al_s(\mu)}{\pi}\cdot \frac{2}{3} \left(
\frac{\pi^2}{3}+1\right) +\left(\frac{\al_s(\mu)}{\pi}\right)^2 c_2(\mu)
 \, .
\eeq
with the recently calculated
3-loop coefficent
 \cite{tarasov} $c_2$
($c_2=-14.59$ for $\mu =m_t$ and 6 flavors).
It reduces the scale
dependence of $\drqcd$ significantly.
The complete
 $O(\al\al_s)$ corrections to the self energies
 beyond the $m_t^2$ approximation are available from
 perturbative  calculations
\cite{qcd} and by means of dispersion relations \cite{dispersion1}.
Quite recently, also non-leading terms to $\Delta r$ of $O(\al_s^2)$
have become available \cite{cks}.
\item
With the quantity $(\Dr)_{remainder}$ in the denominator
non-leading higher order terms
containing mass singularities of the type $\al^2\log(M_Z/m_f)$
from light fermions
are also incorporated \cite{nonleading}.
\end{itemize}

\smallskip \noi
 The quantity $\Dr$ in Eq.\ (10)
$$
\Dr \,=\, 1\,-\,\frac{\pi\al}{\sqrt{2}\Gmu} \, \frac{1}
     {M_W^2 \left(1-\frac{M_W^2}{M_Z^2} \right) } \, .
$$   is
experimentally  determined by $M_Z$ and
$M_W$.
Theoretically, it is computed from $M_Z,\Gmu,\al$
after specifying the masses $M_H,m_t$.
The theoretical prediction for $\Dr$
is displayed in Figure 2.
For comparison with data, the experimental $1\sigma$ limits
from the direct measurements of $M_Z$ at LEP and
$M_W$ in $p\bar{p}$ are indicated.

\begin{figure} % fig 2
\vspace*{3.25in}
% next line was used to print actual photo, commented out here.
% your syntax will probably differ.
% \hbox to\hsize{\hfill\special{ps: epsfile photo.eps}\kern3in\hfill}
\caption{ $\Dr$ as a function of the top mass for
          $M_H=60$ and  $1000$ GeV.
          $1\sigma$ bounds from $M_Z$ and $s_W^2$:
          horizontal band from $p\bar{p}$, $\bullet$ from $\nu N$.}
\end{figure}

The quantity $s_W^2$ resp.\  the ratio $M_W/M_Z$
can indirectly be measured in deep-inelastic neutrino scattering,
in particular in the
NC/CC neutrino         cross section ratio for isoscalar targets.
%The recent CCFR result \cite{bodek}
%  $$ s_W^2 =  0.2222 \pm 0.0057 $$
%combined with the CDHS and
%CHARM results \cite{neutrino} yields the world average \cite{bodek}
The present world average from CCFR, CDHS and CHARM results
\cite{neutrino}
  $$ s_W^2 =  0.2253 \pm 0.0047  $$
is fully consistent with the direct vector boson mass measurements
and with the standard theory.

%\subsection{$Z$ boson observables}
%\vglue 0.4cm
\vglue 0.3cm
\leftline{\twelveit 1.3. $Z$ boson observables}
%\vglue 0.4cm
\vglue 1pt
Measurements
of the $Z$ line shape in $\epmf$
\beq
\sigma(s)   =  \sigma_0\, \frac{s \G_Z^2}
 {\mid s-\mz+i\frac{s}{M_Z^2}\G_Z\mid ^2}
 + \sigma_{\g Z} + \sigma_{\g} \, ,
  \;\;\;\;\;
\sigma_0 = \frac{12\pi}{\mz}\cdot\frac{\G_e\G_f}{\G_Z^2}
\eeq
(with small photon exchange and interference terms)
yield the parameters
$M_Z,\, \G_Z$,   and the partial widths $\G_f$ or the peak
cross section  $\sigma_0$.
Whereas $M_Z$ is used as a precise input parameter, together
with $\al$ and $\Gmu$, the width and partial widths allow
comparisons with the predictions of the Standard Model.
The predictions for the partial widths
as well as for the asymmetries
can conveniently be calculated in terms of effective neutral
current coupling constants for the various fermions.

\paragraph{\it Effective $Z$ boson couplings:}

The effective couplings follow
from the set of 1-loop diagrams
without virtual photons,
the non-QED  or weak  corrections.
These weak corrections
can conveniently be written
in terms of fermion-dependent overall normalizations
$\rho_f$ and effective mixing angles $s_f^2$
in the NC vertices \cite{formfactors}:
\bea
 J_{\nu}^{NC} & = & \left( \sqrt{2}\Gmu\mz \rho_f \right)^{1/2}
\left[ (I_3^f-2Q_fs_f^2)\ganu-I_3^f\ganu\gafi \right] \nn\\
  & = & \left( \sqrt{2}\Gmu\mz \right)^{1/2} \,
  [g_V^f \,\ganu -  g_A^f \,\ganu\gafi]  \, .
\eea
%The complete expressions for $\rho_f,\kappa_f$ can be found in$^{27}$.
%Up to small terms negligible at the $Z$
%peak, they correspond to those of Bardin et al.$^{28}$.
$\rho_f$ and $s_f^2$ contain  universal
parts     (i.e.\ independent of the fermion species) and
non-universal parts which explicitly depend on the type of the
external fermions.
In their leading terms, incorporating also the next order,
the parameters are  given by
\beq
\rho_f  =  \frac{1}{1-\drb} + \cdots , \;\;\;
 s_f^2  = s_W^2 + c_W^2\,\drb + \cdots
\eeq
with $\drb$ from Eq.\ (11).

\smallskip
For the $b$ quark, also the non-universal parts have a strong
dependence on $m_t$ resulting from virtual top quarks in the
vertex corrections. The difference between the $d$ and $b$
couplings can be parametrized in the following way
\beq
  \rho_b = \rho_d (1+\tau)^2, \;\;\;\;
  s^2_b = s^2_d (1+\tau)^{-1}
\eeq
with the quantity
$$
 \tau = \Delta\tau^{(1)}
      + \Delta\tau^{(2)}
      + \Delta\tau^{(\al_s)}
$$
calculated perturbatively, at the present level comprising:
the complete 1-loop order term \cite{vertex}
\beq
\Delta\tau^{(1)} = -2 x_t - \frac{\Gmu\mz}{6\pi^2\sqrt{2}}
 (c_W^2+1)\log\frac{m_t}{M_W} + \cdots , \;\;\;\;
 x_t = \frac{\Gmu m_t^2}{8\pi\sqrt{2}} \, ;
\eeq
 the leading
electroweak 2-loop contribution of $O(\Gmu^2 m_t^4)$
\cite{barbieri,dhl}
\beq
\Delta\tau^{(2)} = -2\, x_t^2 \, \tau^{(2)} \, ,
\eeq
where
 $\tau^{(2)}$ is a function of $M_H/m_t$
with
 $\tau^{(2)} = 9-\pi^2/3$ for $M_H \ll m_t$;
the QCD corrections to the leading term of $O(\al_s\Gmu m_t^2)$
\cite{jeg}
\beq
\Delta\tau^{(\al_s)} =  2\, x_t \cdot \frac{\al_s}{\pi}
 \cdot \frac{\pi^2}{3} \, ,
\eeq
and the $O(\al_s)$ correction to the $\log m_t/M_W$ term in (17),
with a numerically very small coefficient \cite{log}.

\smallskip
\paragraph{\it Asymmetries and mixing angles:}

The effective mixing angles are of particular interest since
they determine the on-resonance asymmetries via the combinations
   \beq
    A_f = \frac{2g_V^f g_A^f}{(g_V^f)^2+(g_A^f)^2}  \, .
\eeq
Measurements of the \ass hence are measurements of
the ratios
\beq
  g_V^f/g_A^f = 1 - 2 Q_f s_f^2
\eeq
or the effective mixing angles, respectively.

\smallskip
\paragraph{\it $Z$ width and partial widths:}

The total
$Z$ width $\Gamma_Z$ can be calculated
essentially as the sum over the fermionic partial decay widths
\beq
 \Gamma_Z = \sum_f \, \Gamma_f + \cdots , \;\;\;\;
 \Gamma_f = \Gamma  (Z\ra f\bar{f})
\eeq
The dots indicate other decay channels which, however,
are not significant.
 The fermionic partial
widths,
 when
expressed in terms of the effective coupling constants
read up to 2nd order in the (light) fermion masses:
\bea
\Gamma_f
  & = & \G_0
 \, \left[
     (g_V^f)^2  +
     (g_A^f)^2 \left(1-\frac{6m_f^2}{\mz}\right)
                           \right]
 \cdot   (1+ Q_f^2\, \frac{3\al}{4\pi} )
     \, +\, \Delta\G^f_{QCD} \nn
\eea
with
$$
\G_0 \, =\,
  N_C^f\,\frac{\sqrt{2}\Gmu M_Z^3}{12\pi},
 \;\;\;\; N_C^f = 1
 \mbox{ (leptons)}, \;\; = 3 \mbox{ (quarks)}.
$$
The QCD correction for the light quarks
with $m_q\simeq 0$ is given by
\beq
 \Delta\G^f_{QCD}\, =\, \G_0
  \left[ (g_V^f) ^2+ (g_A^f)^2 \right]
 \cdot K_{QCD}
\eeq
with \cite{qcdq}
\bea
K_{QCD}  & = &   \frac{\al_s}{\pi} +1.41 \left(
  \frac{\al_s}{\pi}\right)^2 -12.8 \left(
  \frac{\al_s}{\pi}\right)^3 \,  .
\eea
For $b$ quarks
the QCD corrections are different due to  finite $b$ mass terms
and to top quark dependent 2-loop diagrams
 for the axial part:
\bea
 \Delta\G_{QCD}^b & = &
 \Delta\G_{QCD}^d \, +\, \G_0 \left[
           (g_V^b)^2  \, R_V \,+\,
           (g_A^b)^2 \, R_A  \right]
\eea
The coefficients in the perturbative expansions
$$
 R_V = c_1^V \aspi + c_2^V (\aspi)^2 + c_3^V (\aspi)^3 + \cdots,
 \;\;\;
 R_A =  c_1^A \aspi + c_2^A (\aspi)^2 + \cdots \,
$$
depending on $m_b$ and $m_t$,
are calculated up to third order
 in the vector and up to second order
in the axial part \cite{qcdb}.

\smallskip
\paragraph{\it \sm predictions versus data:}

In table 1
the \sm predictions for $Z$ pole observables   are
put together. The first error corresponds to
the variation of $m_t$ in the observed range (2) and $ 60 < M_H < 1000$ GeV.
The second error is the hadronic
uncertainty from $\al_s=0.123\pm 0.006$, as measured
by QCD observables at the $Z$ \cite{alfas}.
 The recent combined LEP results on the $Z$ resonance
parameters \cite{LEP}, under the assumption of lepton universality,
are also shown in table 1, together with $s^2_e$ from
the left-right asymmetry at the SLC \cite{sld}.

The value for the leptonic mixing angle from the left-right asymmetry
$A_{LR}$ has become closer to the LEP result, but due to its smaller
error the deviation is still more than $2\sigma$. One has to keep in
mind, however, that the agreement within the individually determined
values for $s^2_{\ell}$ is much better (see e.g. C. Baltay, these
proceedings). Averaging the results from LEP and SLC yields
$$ s^2_{\ell} = 0.2313 \pm 0.0003 \, . $$

A significant deviation from the \sm prediction is still present in the
quantity $R_b = \Gamma_b/\Gamma_{had}$. The ratio $R_c$ is not precise
enough to claim a deviation from the Standard Model.
\begin{table}
\bc
\caption{LEP results and \sm predictions for the $Z$ parameters.}
 \btab{| l | l | r | }
\hline
 observable & LEP 1995 & \sm prediction \\
\hline
\hline
$M_Z$ (GeV) & $91.1887\pm0.0022$ &  input \\
\hline
$\Gamma_Z$ (GeV) & $2.4971\pm 0.0032$ & $2.4976 \pm 0.0077\pm 0.0033$ \\
%\hline
%$\Gamma_{had}$ (GeV) & $1.740\pm 0.008$ &
% $1.736\pm 0.008 \pm 0.007$ \\
%\hline
%$\Gamma_e$ (MeV) & $83.2\pm 0.4$ & $83.7\pm 0.4 $ \\
\hline
$\sigma_0^{had}$ (nb) & $41.492\pm 0.081$ & $41.457\pm0.011\pm0.076$ \\
\hline
 $\G_{had}/\G_e$ & $20.800\pm 0.035 $ & $20.771\pm 0.019\pm 0.038$ \\
%\hline
%$\Gamma_e$ (MeV) & $83.82\pm 0.27$ & $83.7\pm 0.4 $ \\
\hline
$\Gamma_{inv}$ (MeV) & $499.5\pm 2.7$ & $501.6\pm 1.1$ \\
\hline
$\G_b/\G_{had}$  & $0.2204\pm 0.0020$ & $0.2155\pm 0.0004$ \\
\hline
$\G_c/\G_{had}$  & $0.1606\pm 0.0095$ & $0.1713\pm 0.0002$ \\
  \hline
$\rho_{\ell}$ & $1.0049\pm 0.0017$ & $1.0050\pm 0.0023$ \\
\hline
$s^2_{\ell}$ & $0.2318\pm 0.0004$ & $0.2317\pm 0.0012$ \\
\hline
$s^2_e (A_{LR})$ & $0.2305\pm 0.0005$ & $0.2317\pm 0.0012$   \\
        &  (SLC result)               &                    \\
\hline
\etab
\ec
\end{table}

\smallskip
\paragraph{\it \sm fits:}

Assuming the validity of the \sm a global fit to all electroweak
LEP results
constrains the parameters $m_t,\al_s$ as follows: \cite{LEP}
\beq
    m_t = 176\pm 10^{+17}_{-19}\, \gv, \;\;\
    \al_s = 0.125 \pm 0.004 \pm 0.002
\eeq
with $M_H= 300$ GeV for the central value.
The second error is from the variation of $M_H$
between 60 GeV and 1 TeV.
The fit results include the
uncertainties of the \sm calculations to be discussed in the
next subsection.
The parameter range in Eq.\ (27) predicts a value
for the $W$ mass  via Eq.\ (8,10)
$$ M_W = 80.32\pm 0.06\pm 0.01 \, \gv \, , $$
in best agreement with the direct measurement, Eq.\ (3),
 but with a sizeably smaller  error.
Simultaneously, the result (26)
is a consistency check of the QCD part
of the full  \sm: the value of
$\al_s$ at the $Z$ peak, measured from others than electroweak
observables, is         \cite{alfas}
$\al_s=0.123 \pm 0.006$.

\smallskip
\paragraph{\it Low energy results:}

The  measurement of the mixing angle in neutrino-$e$ scattering
by the CHARM II Collaboration yields \cite{beyer}
\beq
 \sin^2\! \theta_{\! \nu\! e} = 0.2324\pm 0.0083 \, .
\eeq
 This value coincides with the LEP result
on $s^2_{\ell}$, table 1, as expected by the theory. The major
sources of a potential
difference: the different scales and the neutrino
charge radius, largely cancel each other by numerical coincidence
\cite{cradius}.

\smallskip
The results from deep inelastic $\nu$ scattering have already been
discussed in the context of $M_W$. Including the information from
CDHS, CHARM, CCFR, and  with $M_W$ from $\bar{p}p$
modifies the fit result only marginally \cite{LEP}:
\beq
    m_t = 174\pm 9^{+17}_{-19}\, \gv, \;\;\;
    \al_s = 0.126 \pm 0.004 \pm 0.002\, .
\eeq
Incorporating the SLC result on $A_{LR}$ yields \cite{LEP}
\beq
    m_t = 179\pm 9^{+17}_{-19}\, \gv, \;\;\;
    \al_s = 0.125 \pm 0.004 \pm 0.002\, .
\eeq

\medskip
The main Higgs dependence of the electroweak predictions is only
logarithmic in the Higgs mass. Hence, the sensitivity of the data
to $M_H$ is not very pronounced. Using the Tevatron value for $m_t$ as
an additional experimental constraint, the electroweak fit to all data
yields $M_H < 900$ GeV with approximately 95\% C.L. \cite{LEP}.
Similar results have been obtained in \cite{warsaw}.

\smallskip
A  fit to $m_t$ leaving $M_H$ free
 yields a slightly lower range
\cite{jellis}
$ m_t =155 \pm 15$ GeV.
 The reason is the
theoretical correlation between $m_t$ and $M_H$ together with
the lower $\chi^2$ values in the fits for smaller Higgs masses.

\section{Status of the Standard Model predictions}
%\vglue 0.4cm
%\vglue 0.3cm
%\leftline{\twelveit 1.4 Status of the Standard Model predictions}
%\vglue 0.4cm
%\vglue 1pt
 For a discussion of the theoretical reliability
of the \sm predictions one has to consider the various sources
contributing to their
uncertainties:

The experimental error propagating into the hadronic contribution
of $\al(\mz)$, Eq.\ (6), leads to
$\delta M_W = 13$ MeV in the $W$ mass prediction, and
$\delta\sin^2\theta = 0.00023$ common to all of the mixing
angles, which matches with the future experimental precision.

The uncertainties from the QCD contributions,
 besides the 3 MeV in the
hadronic $Z$ width, can essentially be traced back to
those in the top quark loops for the $\rho$-parameter.
They  can be combined into the following errors
\cite{kniehl95}, which have improved due to the recently available
3-loop results:

$$
 \delta(\dro) \simeq 1.5\cdot 10^{-4},   \;
 \delta s^2_{\ell} \simeq 0.0001
$$
for $m_t = 174$ GeV, and slightly larger for heavier top.

The size of unknown higher order contributions can be estimated
by different treatments of non-leading terms
of higher order in the implementation of radiative corrections in
electroweak observables (`options')
and by investigations of the scheme dependence.
Explicit comparisons between the results of 5 different computer codes
based on  on-shell and $\ms$ calculations
for the $Z$ resonance observables are documented in the ``Electroweak
Working Group Report'' \cite{ewgr} of the recent
``Reports of the Working
Group on Precision Calculations for the $Z$ Resonance''
\cite{yellow}. The typical size of the genuine electroweak uncertainties
is of the order 0.1\%.
For the leptonic mixing angle, the most severe case, one finds
$$\delta s^2_{\ell} \simeq 1.5 \cdot 10^{-4} \, , $$
which is again of the same order as the experimental precision.
Improvements require systematic 2-loop calculations.
As an example, the leptonic mixing angle is displayed in Figure 3.

\smallskip
Low angle
Bhabha scattering for a luminosity measurement at 0.1\% accuracy
still requires more theoretical effort. For a description of the present
status see the contributions by Jadach et al.\ and other authors
in \cite{yellow}.

\section{Virtual New Physics}
%\vspace*{-0.7cm}
%\subsection{Parametrization of precision observables}
%\vspace*{-0.35cm}

The parametrization of the radiative corrections originating
 from the vector
boson self-energies in terms of the static $\rho$-parameter
$\dro(0) \equiv \eps_1$
and two other combinations of self-energies, $\eps_2$ and $\eps_3$,
\cite{epsilon}
allows a generalization of the analysis of the electroweak data
which accommodates extensions of the minimal model
affecting only the vector boson self-energies.
There is a wide literature \cite{pt} in this field  with various
conventions.

Phenomenologically, the $\eps_i$ are parameters which
can be determined experimentally from
the normalization of the $Z$ couplings and
  the effective mixing angle by
(the residual corrections not from self-energies are dropped)
\beq
 \rho_f = \dro(0) + \mz \Pi^{'\,ZZ}(\mz)
           + \cdots, \;\;\;
    s_f^2 = (1+\Delta\kappa') s_0^2 \, + \cdots
\eeq
with $s_0^2$ from Eq.\ (26)  and
\beq
\Delta\kappa' = -\frac{c_0^2}{c_0^2-s_0^2} \dro(0)
      + \frac{\eps_3}{c_0^2-s_0^2} \, ,
\eeq
the quantity $\Delta r$ in the $M_W$-$M_Z$ correlation:
\beq
 \Delta r = \Delta\al - \frac{c_0^2}{s_0^2}\dro(0)
 +\frac{c_0^2-s_0^2}{s_0^2}\eps_2 + 2\eps_3  \, .
\eeq
The $\eps$ parameters have been redefined
\cite{abc1} into $\eps_{N1,N2,N3}$ by including also the
$v$ and $a$ vertex corrections for leptons, together with a
4th quantity $\eps_b$ to parametrize specific non-universal
left handed contributions to the $Zbb$ vertex via
\beq
 g_A^b =g_A^d(1+\eps_b), \;\;\;
 g_V^b/g_A^b = (1-\frac{4}{3}s^2_d+\eps_b)\, (1+\eps_b)^{-1} \, .
\eeq
Figure 3 shows the results from a global data analysis in terms of
the $1\sigma$ contours.
The level of consistency with the \sm is visualized
by the \sm predictions displayed in terms of the lines
 with $m_t,M_H$ as input quantities.
The displacement of the $\eps_b$-contours corresponds to the difference
between the \sm prediction and the experimental result for $R_b$
(see table 1).
Among the alternative mechanisms of electroweak symmetry breaking,
most versions of technicolor models are disfavored
by the data \cite{jellis1,technicolor}.

\begin{figure} % fig 3
\vspace*{12cm}
%\caption{    $\Delta\chi^2=1$ contours for the $\eps$ parameters
\caption{    $1\sigma$ contours for the $\eps$ parameters
             and the \sm predictions, from ref 43}
\end{figure}

\smallskip
Attempts to attribute the observed difference in $R_b$ to new physics
in the $Zbb$ vertex have to obey the constraints from
the other observables, in particular from $R_h = \Gamma_{had}/\Gamma_e$
and $\Gamma_Z$. In this way, a  value for $\alpha_s$ is obtained
which is about $1\sigma$ lower than the one from the \sm fit
 \cite{schild,warsaw,garcia}.

\medskip
A current example of new physics with also extra vertex contributions
is the \sm with two Higgs doublets. The charged Higgs bosons diminish
the value of $R_b$ even more and hence are strongly constrained,
clearly disfavored for small values of $\tan\beta=v_2/v_1$
\cite{cornet}. Also the neutral sector of the general 2-doublet model
turns out to be severely constrained \cite{cornet}.

\medskip
A special discussion deserves the minimal supersymmetric standard
(MSSM)
model as the most predictive framework beyond the minimal model.
Its structure allows a similarly complete calculation of
the electroweak precision observables
as in the Standard Model in terms of one Higgs mass
(usually taken as $M_A$) and $\tan\beta$, together with the set of
SUSY soft breaking parameters fixing the chargino/neutralino and
scalar fermion sectors.
It has been known since quite some time
\cite{higgs,susy0,abc2,susy1,susy2,susy3,susy4,susy5}
that light non-standard
Higgs bosons as well as light stop and charginos, all around 50 GeV or
little higher,
yield larger values for the ratio $R_b$ and thus diminishing the observed
difference. Complete 1-loop calculations are meanwhile available for
$\Delta r$ \cite{susydelr} and for the $Z$ boson observables
\cite{susy3,susy4,susy5}.

\smallskip
In Figure 4 the range of the theoretical predictions for the various
observables are displayed for the \sm and the MSSM ($\al_s = 0.123)$.
 In the minimal model,
$M_H$ is varied as usual between 60 GeV and 1 TeV (dashed curves).
The MSSM range (between the full lines) are obtained for $\tan\beta$
between 1.1 and 70, and $60 < M_A < 1000$ GeV, all other SUSY particles
taken with masses obeying the present bounds from direct searches.
The shaded areas denote the experimental 1$\sigma$ bounds.
The prefered parameter domain yielding the optimum agreement with the
data comprises low values for stop, chargino and
$M_h, M_A$, close to present lower limits.
This is made more explicit by a global fit to the precision data
 performed in \cite{susy4}.
Simultaneously, $\al_s$ turns out to be closer  to the world average
0.118
\cite{garcia,susy4}
 (mainly from $\G_Z$
and $R_{had}$).

\section{Conclusions}

The agreement
of the experimental high and low energy precision
data with the \sm predictions has shown that the \sm works as a fully
fledged quantum field theory.
A great success of the \sm is the experimentally observed
top mass range which coincides in an
impressive way with
the indirect determination through loop effects from precision data.

The steadily increasing accuracy of the data starts to exhibit
also sensitivity to the Higgs mass,
 although still marginally.

Still not understood at present is the deviation
from the theoretical expectation
 observed in the
measurement of $R_b$. Among the possible extensions of the minimal model,
supersymmetry seems to be a favorite candidate which can accomodate also
a large $R_b$ value without contradicting the other data as long as $m_t$
is not too high and non-standard particles in the discovery range of
LEP II are around.

\bigskip
{\bf Acknowledgement:}

 I want to thank G. Altarelli,
 A. Dabelstein, R. Ehret, M. Gr\"unewald and D. Schaile
for valuable information and helpful discussions. \\[1cm]

 \end{document}